\documentclass[sigconf]{acmart}
\usepackage{algorithm}
\usepackage{algpseudocode}
\usepackage{pifont}
\usepackage{multirow}
\usepackage{siunitx}
\usepackage{makecell}

\graphicspath{{./arxiv_fig/}}
\AtBeginDocument{%
  }
\renewcommand\footnotetextcopyrightpermission[1]{}
\acmConference{}{}{}
\acmYear{}
\acmDOI{}
\acmISBN{}
\begin{document}
\makeatletter
\let\ps@acm\ps@plain 
\def\@runninghead{} 
\makeatother
\fancyhf{}
\renewcommand{\headrulewidth}{0pt}
\title{NVLLM: A 3D NAND-Centric Architecture Enabling Edge on-Device LLM Inference}
\author{
    Mingbo Hao,
    Changwei Yan,
    Haoyu Cui,
    Zhihao Yan,
    Yizhi Ding,
    Zhangrui Qian,
    Weiwei Shan$^{\scriptscriptstyle *}$
}
\thanks{Emails: mingbohao@seu.edu.cn (Mingbo Hao); wwshan@seu.edu.cn (Weiwei Shan, corresponding author).}
\thanks{ 
\par\vspace{-2pt}\noindent\rule{\linewidth}{0.4pt}\par\vspace{2pt}%
Mingbo Hao is a Ph.D. student at Southeast University, pioneering 3D NAND-centric computing architectures for large language model inference acceleration.
This document is an author-prepared preprint version of the manuscript, with publication-specific formatting elements removed for improved readability. 
The work presents a system-level exploration of 3D NAND-centric LLM inference acceleration under heterogeneous memory hierarchies, with NVLLM serving as an initial system instantiation of this research direction. This version incorporates minor refinements from the conference version while preserving the core design, evaluation methodology, and conclusions. It is released for academic dissemination and discussion.
\vspace{3.5em}
}

\affiliation{
  \institution{School of Integrated Circuits, Southeast University, Nanjing, China}
  \country{}
}

\begin{abstract}

The rapid growth of LLMs demands high-throughput, memory-capacity-intensive inference on resource-constrained edge devices, where single-batch decoding remains fundamentally memory-bound. Existing out-of-core GPU-based and SSD-like accelerators are limited by DRAM-bound weight movement and inefficient storage access granularity. We present NVLLM, a 3D NAND-centric inference architecture that offloads feed-forward network (FFN) computation into the Flash while executing attention on lightweight CMOS logic with external DRAM. Through wafer-to-wafer stacking, NVLLM tightly integrates multi-plane 3D NAND with compute pipelines, error correction code (ECC) units, and buffers, enabling page-level FFN weight access without DRAM traversal. All GEMM/GEMV operations are decomposed into dot-product primitives executed by out-of-order PE lanes, operating directly on raw NAND reads with integrated ECC. Attention weights remain in DRAM, and a KV-cache-aware scheduler sustains throughput as the context length grows. Evaluated on OPT and LLaMA models with up to 30B parameters, NVLLM achieves a 16.7$\times$--37.9$\times$ speedup over A800-based out-of-core inference and up to 4.7$\times$ speedup over SSD-like designs, with only 2.7\% CMOS area overhead.

\end{abstract}




\maketitle

\section{Introduction}

\begin{figure}[t]
    \centering
    \includegraphics[width=\linewidth]{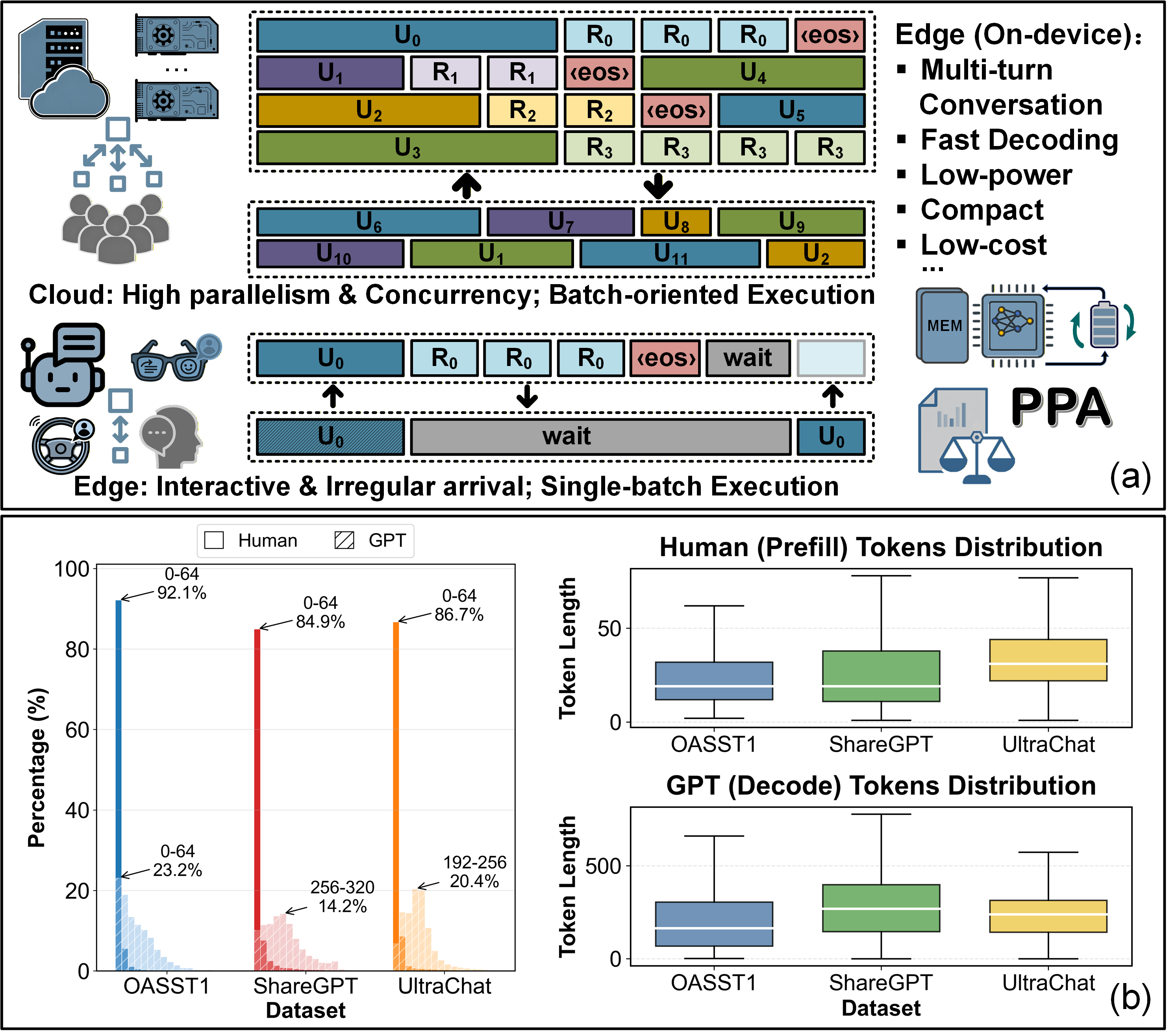}
    \vspace{-12pt} 
    \caption{(a) Comparison of edge and cloud LLM inference devices and their execution characteristics.
    (b) Analysis of the token-distribution patterns in human--LLM interactive conversations.}
    \label{fig:fig1}
\end{figure}

Deploying large language models (LLMs) on edge devices is increasingly appealing for privacy, latency, and offline availability~\cite{isca26_dialogue_1, isca26_LLM_edge_1, dabian_ppt_survey}. However, achieving accurate and efficient on-device inference remains challenging due to strict constraints on memory capacity, bandwidth, and power. Unlike cloud servers, edge devices cannot exploit large-batch parallelism; interactive requests arrive irregularly, and decoding predominantly proceeds in a single-token, single-batch manner~\cite{isca26_dialogue_1}. As shown in Fig.~\ref{fig:fig1}(a), cloud workloads naturally form high arithmetic intensity~\cite{new_orca, new_neupim}, whereas edge workloads do not.

To characterize this behavior, Fig.~\ref{fig:fig1}(b) analyzes 6.1M conversation turns from ShareGPT~\cite{ShareGPT_CNEN_90k}, UltraChat~\cite{ding2023enhancing}, and OASST1~\cite{OpenAssistant_OASST1}. User prompts are overwhelmingly short, while model responses exhibit long tails. This ``short-prompt, long-generation'' pattern implies that edge inference is dominated by the decode phase, where each token requires a full forward pass without batch-level parallelism.

\begin{figure}[t]
    \centering
    \includegraphics[width=\linewidth]{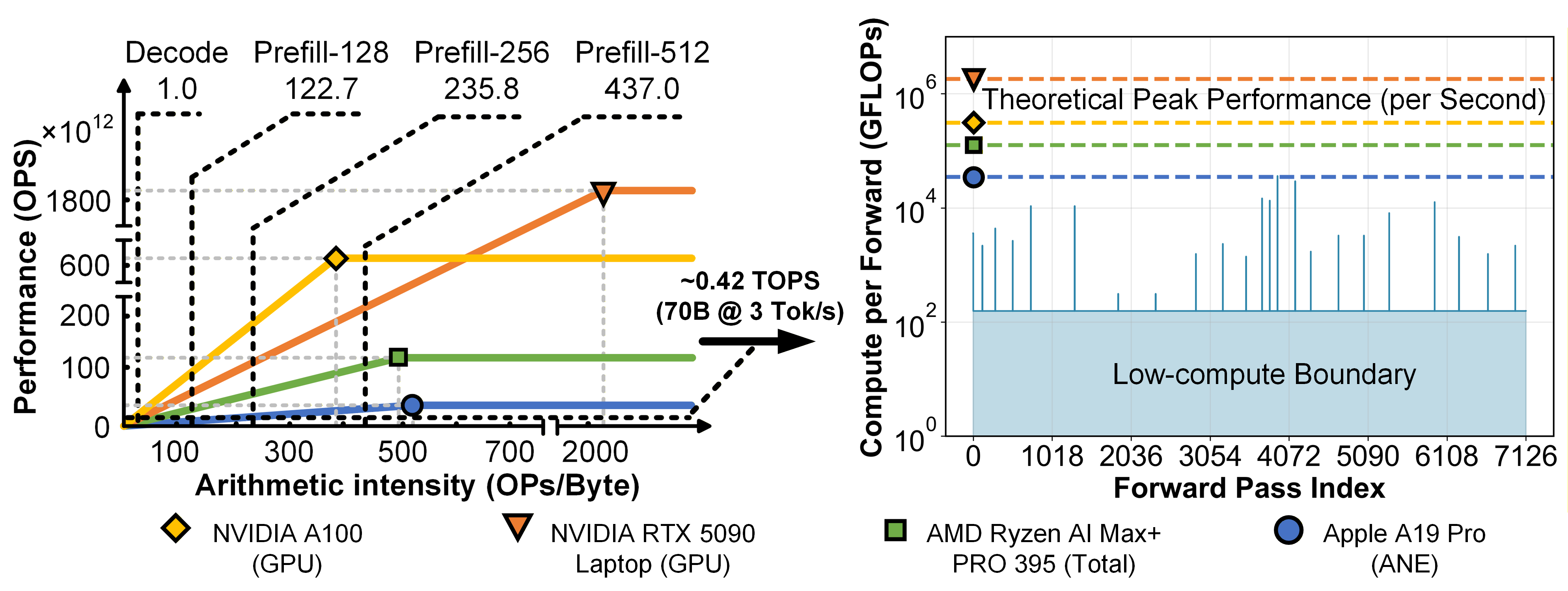}
    \vspace{-13pt} 
    \caption{Under the single-batch inference constraint, the hardware remains heavily memory-bound for the majority of the workload.}
    \label{fig:fig2}
\end{figure}

Fig.~\ref{fig:fig2} shows a roofline analysis across representative hardware platforms~\cite{Apple_Silicon, NVIDIA_RTX50_Laptops, AMD_RyzenAI_395, Nvidia_A100}. Under single-batch, long-generation workloads, performance is heavily memory-bound during both prefill and decode. For instance, maintaining a 3 tokens/s user-perceived rate~\cite{3TokenpreS_1, 3TokenpreS_2} for a 70B model requires only 0.42 TOPS of theoretical compute---comparable to the per-token rate of an RTX 4090 running 6-bit LLaMA3.3-70B~\cite{llamacpp}---yet its peak compute is over 3000$\times$ higher. A multi-turn ShareGPT trace further confirms that decode iterations lie far below hardware compute ceilings but remain bottlenecked by weight loading.

Since batch-level parallelism is fundamentally limited on the edge, datacenter-oriented optimizations~\cite{isca26_Splitwise} lose effectiveness. Edge-side approaches such as speculative decoding~\cite{isca26_touji}, lookahead decoding~\cite{isca26_qianzhan}, early exit~\cite{isca26_zaotui}, and quantization/sparsity~\cite{wangSpAttenEfficientSparse2021} remain limited by DRAM bandwidth. Increasing DRAM bandwidth (e.g., HBM~\cite{isca26_hbm_1}) raises power and cost; insufficient DRAM capacity forces repeated parameter streaming~\cite{sheng2023flexgen}, further increasing latency and energy. Thus, scaling DRAM alone is not a viable solution for single-batch inference.

3D NAND Flash provides an alternative path. Single-token decoding offers substantial compute slack, enabling near-data computation to amortize data movement costs. Wafer-to-wafer (W2W) 3D integration~\cite{koWaferlevel3DIntegration2010, yolandaWaferWaferBonding2022} stacks CMOS beneath or above the NAND array, allowing lightweight compute without sacrificing density. Modern Flash products already adopt such structures~\cite{xtacking_vlsi, TechInsights_Xtacking40_WhatsNew}, combining high density and low cost per bit with the ability to access multi-megabyte feed-forward network (FFN) weights in situ.

However, several challenges arise. Raw NAND reads contain bit errors requiring error correction code (ECC), yet in-Flash computation precedes conventional correction. Flash writes are slow (200--800~$\mu$s program, 3--5~ms erase) and endurance-limited~\cite{ytmc_TemperatureEffectsProgram2025}, constraining write-heavy schemes. Meanwhile, KV-cache growth increases attention cost and can degrade decode throughput if unmanaged.

We propose NVLLM, a \textbf{3D NAND-centric} architecture addressing these challenges. FFN weights are stored in Flash, while attention weights reside in DRAM. GEMM/GEMV operations are decomposed into dot-product primitives executed by an out-of-order in-Flash pipeline. By decoupling error detection from correction, the pipeline operates directly on raw NAND reads, sustaining throughput while tolerating bit errors. Keeping attention weights in DRAM reduces bandwidth pressure and preserves the random-access demands of the attention path. A KV-cache–aware scheduler monitors cache growth and dynamically offloads portions of Q/K/V/O projections to the in-Flash pipeline, maintaining stable per-token throughput under long contexts.

Building on these design principles, NVLLM shows that combining in-Flash FFN execution, DRAM-resident attention weights, and KV-cache–aware scheduling effectively improves token throughput, reduces off-chip memory traffic, and scales to larger models under edge constraints. Evaluated on quantized OPT and LLaMA models up to 30B parameters, NVLLM achieves 16.7$\times$--37.9$\times$ higher token throughput than an out-of-core GPU under the same weight-offloading strategy, and up to 4.7$\times$ higher throughput over SSD-like in-Flash designs, while incurring only 2.7\% CMOS area overhead. The key contributions of this work are as follows:

\begin{itemize}

\item We propose NVLLM, a 3D NAND-centric architecture for edge-side LLM inference. By co-designing compute and storage through wafer-level 3D integration, NVLLM performs FFN computation inside Flash and substantially reduces off-chip memory traffic.

\item We design an in-Flash dot-product engine that sustains high throughput under raw NAND read errors. Decoupling error detection from correction enables out-of-order, error-tolerant vector execution and unlocks an efficient area-power design space.

\item We show that attention and FFN exhibit weak interdependence during inference and exploit this property with a decoupled execution strategy. A KV-cache-aware scheduler further preserves token throughput by adaptively balancing attention workloads in long-context scenarios.

\end{itemize}
\section{Background and Motivation}

\subsection{Large Language Models at the Edge}

Most modern LLMs adopt multi-layer Transformer decoder architectures~\cite{zengFlightLLMEfficientLarge2024}, where each decoder layer consists of an attention module and a FFN. Attention typically employs multi-head attention (MHA), computing queries ($Q$), keys ($K$), and values ($V$) via linear projections, followed by scaled dot-product, causal masking, softmax, and weighted aggregation. FFNs expand the hidden dimension (e.g., $3.5\times$ in LLaMA3-8B) and project it back, accounting for roughly 70\% of total parameters—around $5.3\,\mathrm{GiB}$ under 8-bit quantization. Both attention and FFN maintain input dimensions through residual connections and layer/RMS normalization.

\begin{figure}[t]
  \centering
  \includegraphics[width=7.5cm]{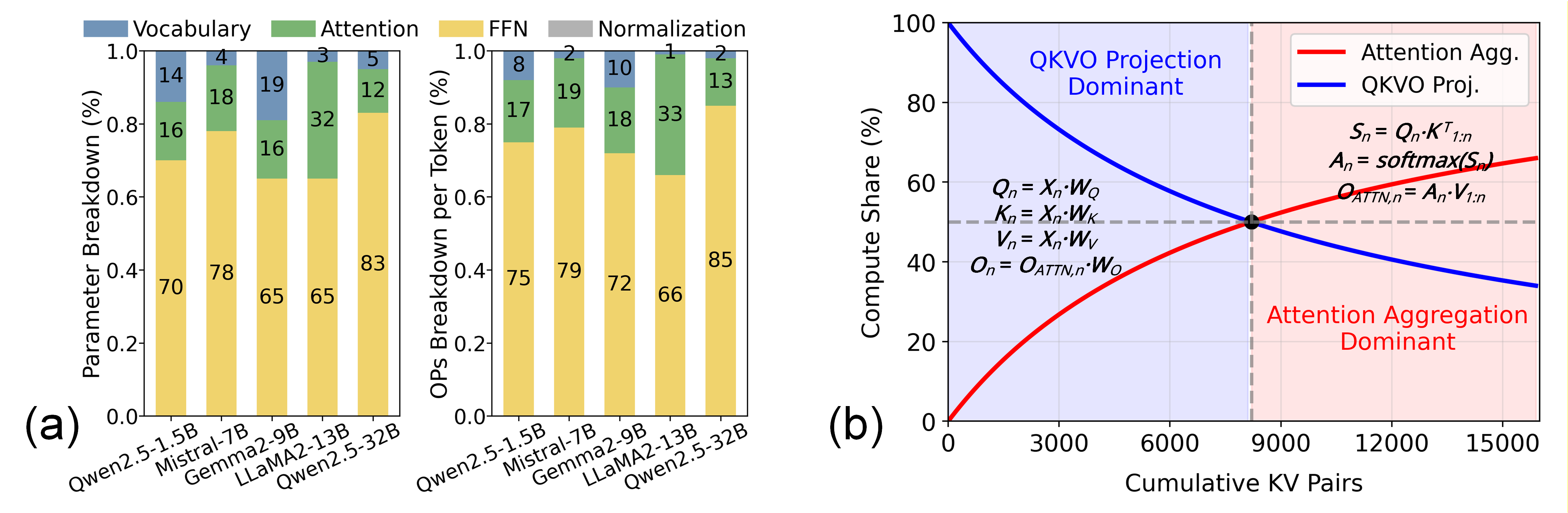}
  \vspace{-12pt} 
  \caption{(a) Breakdown of model parameters and per-token OPs (single batch); Normalization is shown in the legend even if negligible. (b) Attention computation growth with increasing KV-cache size in long-context inference.}
  \label{fig:fig3}
\end{figure}

Edge deployments face fundamentally different workloads than cloud servers~\cite{new_orca}. Interactive single-batch prompts dominate, leaving little room for parallelism and resulting in underutilized compute. Workload analysis (Fig.~\ref{fig:fig3}(a)) confirms that FFNs dominate model parameters, while single-batch decoding dominates runtime compute, making external memory access (EMA) the main energy and throughput bottleneck. Multi-turn interactions increase KV-cache size, shifting attention compute toward aggregation over Q/K/V/O projections (Fig.~\ref{fig:fig3}(b)). These characteristics emphasize the memory- and bandwidth-bound nature of edge LLM inference and highlight the need for architectures tailored to these constraints.

\subsection{3D NAND Flash for Edge LLMs}

Traditional attention-focused accelerators~\cite{new_Sanger, new_spatten, new_Energon} typically store all weights in DRAM, incurring high energy and bandwidth costs (56--69\% of total system power) and up to 1.5$\times$ extra computation. Scaling DRAM alone is costly and often insufficient for single-batch edge inference.  

\begin{figure}[t]
  \centering
  \includegraphics[width=7.6cm]{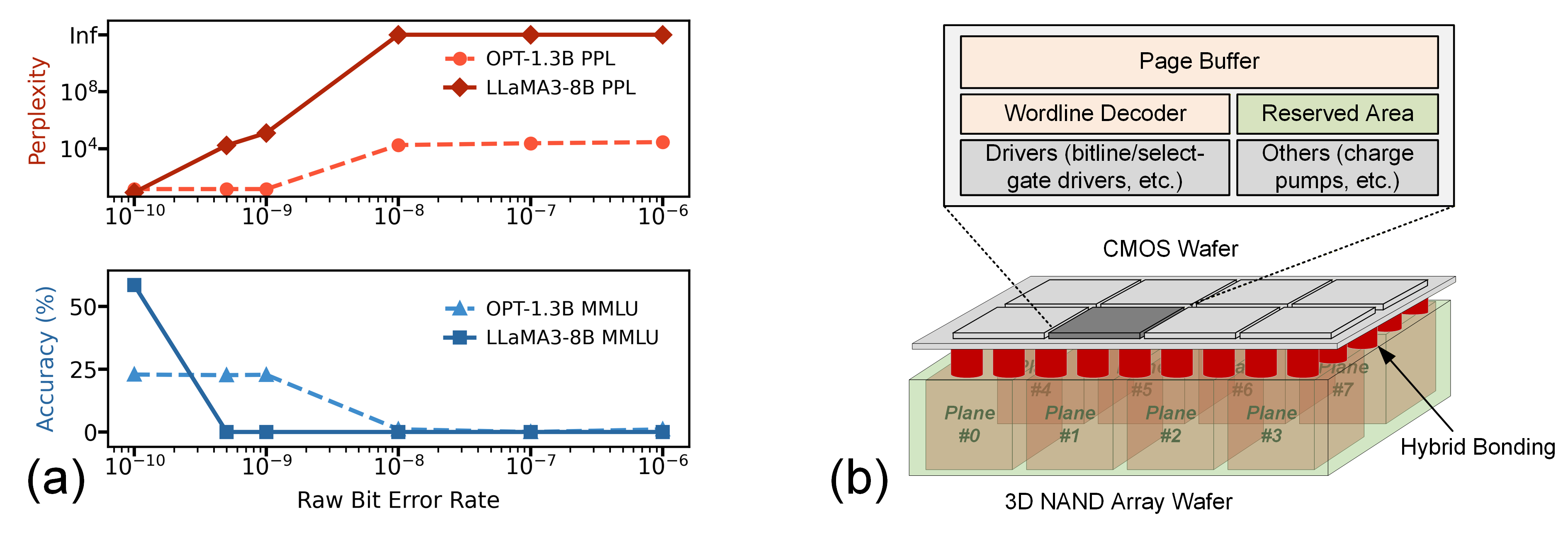}
  \hspace{-0.5cm}
  \vspace{-10pt} 
  \caption{(a) NAND read-induced RBER increases perplexity and lowers accuracy. (b) CMOS area constraints in Flash logic require lightweight, error-tolerant computation pipelines.}
  \label{fig:fig4}
\end{figure}

3D NAND Flash provides an opportunity to offload FFN computation near memory. Single-token decoding offers ample compute slack, while page-level access retrieves multi-gigabyte weights in situ. W2W 3D integration~\cite{huangWaferLevelSystem2021,koWaferlevel3DIntegration2010} allows CMOS logic to reside beneath the NAND array, enabling high-density and area-efficient computation. Challenges include non-negligible raw bit error rate (RBER) in NAND Flash memory, limited write endurance~\cite{ytmc_TemperatureEffectsProgram2025}, and restricted CMOS area for logic~\cite{xtacking_vlsi, micheloniArrayArchitectures3D2017}, as illustrated in Fig.~\ref{fig:fig4}.

Decoupling FFN and attention computation enables efficient in-Flash FFN execution while retaining attention in DRAM, where irregular data dependencies and growing KV caches are better supported. FFN pipelines in Flash can exploit regular, weight-resident access patterns with minimal on-chip buffering, while attention benefits from DRAM bandwidth and algorithmic optimizations. This division improves throughput, reduces off-chip traffic, and scales to longer contexts, motivating a NAND-centric edge architecture for LLM inference.

\begin{figure*}[ht]  
\centering
\includegraphics[width=16cm]{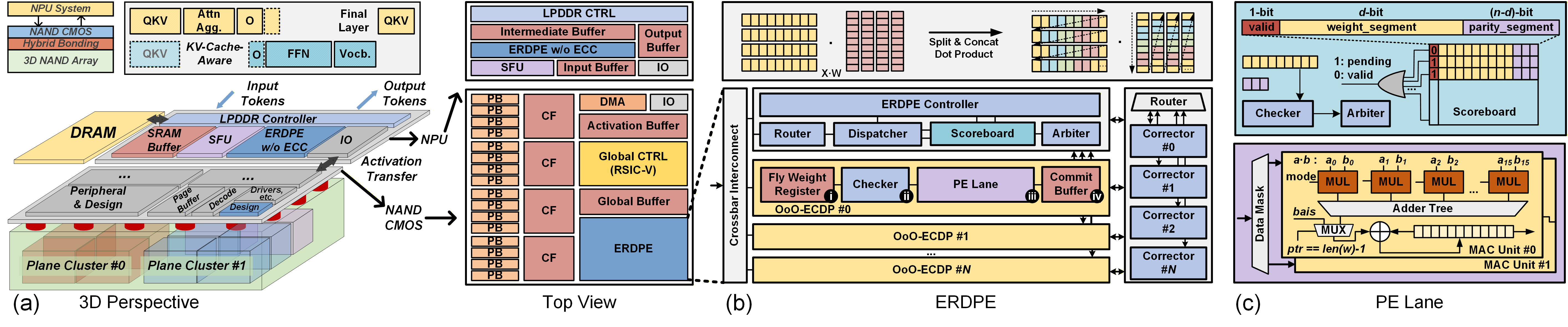}
\vspace{-10pt} 
\caption{Proposed NVLLM architecture.}
\label{fig:fig5}  
\end{figure*}
\section{NVLLM Architecture}

Fig.~\ref{fig:fig5} illustrates the overall architecture of NVLLM, a wafer-to-wafer co-designed system that executes FFN layers directly within 3D NAND Flash, while performing attention and KV-cache operations on the DRAM-side NPU. This division of labor matches the intrinsic characteristics of LLM workloads: FFN layers hold the vast majority of parameters with highly regular, weight-stationary access patterns and minimal activation footprint, whereas attention computation is dominated by cache-resident Q/K/V/O projections and KV-dependent aggregation whose irregular data reuse is best served by DRAM bandwidth.

By combining hybrid-bonded CMOS logic with Flash planes, NVLLM eliminates DRAM round-trips for FFN weights, reduces EMA energy, and frees bandwidth for attention. The architecture is organized around three principles: (i) structured decoupling of attention and FFN compute, (ii) plane–lane co-alignment for weight streaming efficiency, and (iii) error-resilient, out-of-order FFN execution that matches NAND read behavior.

\subsection{Clustered Plane-Lane Co-Design}

3D NAND exposes data at page granularity (4--16\,KiB), but page-level bandwidth from a single plane is insufficient to saturate compute pipelines. NVLLM therefore groups planes into \emph{clusters}, each comprising $k \times k$ planes sharing peripheral circuitry and synchronously returning page-buffer (PB) rows. This arrangement offers two co-design advantages. First, a cluster aggregates the per-plane read bandwidth and amortizes PB/peripheral-circuit overhead, which is particularly beneficial in W2W-bonded 3D NAND, where the CMOS periphery beneath each array footprint is area-constrained. Second, and more importantly, the cluster width is chosen to \emph{match the data consumption width of PE lanes}. Each PE lane processes a segment of $d$ weights per cycle, and a cluster produces $d$ bytes per activation step on average. This alignment eliminates wide multiplexers, avoids cross-lane reshaping, and enables direct streaming from cluster-level FIFOs (CFs) into lane-level segment buffers with a single narrow handshake. Without such co-alignment, implementing a lane-scaled dot-product pipeline within the NAND CMOS wafer would incur prohibitive buffering and routing overhead.

Each cluster thus forms a self-consistent weight delivery unit with three buffering layers: (i) plane-level PBs, (ii) a shallow cluster FIFO that absorbs WL-level read latency, and (iii) per-lane segment buffers that gate data into the compute pipeline. These layers together hide WL RC delay, sense-amplifier variation, and multi-$V_{\text{read}}$ cycles for MLC/TLC cells, enabling deterministic, stall-free weight delivery to FFN compute.

\subsection{Error-Resilient Dot Product Algorithm}

At the heart of in-Flash FFN execution is the error-resilient dot-product engine (ERDPE), which performs mixed-precision (BF16 or INT8) dot products directly on raw NAND reads. To address bit errors inherent to in-Flash computing, NVLLM decouples error detection from correction and introduces a lightweight out-of-order (OoO) dot-product pipeline that implements an error-corrected dot-product (ECDP) in an out-of-order manner (OoO-ECDP).

To realize this execution model in hardware, NVLLM adopts algorithm~\ref{alg:OoO-ECDP} that decomposes each weight column into multiple independent segments and schedules them based on error-check readiness. This algorithm provides a hardware-friendly way to overlap error detection, multi-cycle correction, and MAC execution without requiring the correction units to finish beforehand.
\begin{algorithm}
  \caption{Out-of-Order Dot Product for Error-Resilient}
  \label{alg:OoO-ECDP}
  \begin{algorithmic}[1]
  \Statex \hspace{-2em} \textbf{Input:} 
  Weight vector $\mathbf{w} \in \mathbb{R}^{h \times 1}$,
  Activation vector $\mathbf{a} \in \mathbb{R}^{1 \times h}$,
  Segment factor $d$,
  Error-correcting code $L(n, d)$, 
  Parity matrix $\mathbf{p} \in \mathbb{R}^{(n-d) \times (h/d)}$ ,
  \Statex \hspace{-2em} \textbf{Output:} Final dot product result $s$
  
  \State Initialize accumulator $s \gets 0$, scoreboard $B \gets \emptyset$
  \State Weight vector pointer $ptr \gets 0$ 
  
  \While{$ptr < \text{len}(\mathbf{w})$}
      \State $\mathbf{w}_d \gets \mathbf{w}[ptr:ptr+d-1]$ \Comment{Current weight segment}
      \State $\mathbf{p}_{n-d} \gets \mathbf{p}[0:n-d, ptr]$ \Comment{Parity segment}
      \State $\mathbf{v} \gets \text{Concat}(\mathbf{w}_d, \mathbf{p}_{n-d})$
      
      \If{$\text{Checker}(\mathbf{v}, L(n,d))$}
          \State $s \gets s + \mathbf{w}_d \cdot \mathbf{a}[ptr:ptr+d-1]$
      \Else \Comment{Non-blocking}
          \State $B \gets B \cup \{ptr\}$
          \State $\mathbf{w}' \gets \text{Corrector}(\mathbf{w}_d, \mathbf{p}_{n-d})$
          \State $\mathbf{w}[ptr:ptr+d-1] \gets \mathbf{w}'$
      \EndIf
      \State $ptr \gets ptr + d$
  \EndWhile
  
  \If{$B \neq \emptyset$}
      \For{$idx \in B$}
          \State $\mathbf{w}_d \gets \mathbf{w}[idx:idx+d-1]$
          \State $s \gets s + \mathbf{w}_d \cdot \mathbf{a}[idx:idx+d-1]$
      \EndFor
  \EndIf
  
  \State \Return $s$
  \end{algorithmic}
\end{algorithm}
\subsection{ERDPE Microarchitecture}

ERDPE consists of multiple OoO-ECDPs, each containing PE lanes that execute segment-level dot products~\ref{fig:fig5}(b--c). For each weight segment, a lane first performs an inline error check. If no error is detected, the segment proceeds directly to MAC accumulation. Otherwise, the segment is forwarded to a shared multi-cycle corrector hub, while the lane continues processing other ready segments. Corrected segments returned from the hub update the partial sums tracked by the scoreboard.

Each OoO-ECDP implements the OoO dot-product algorithm in hardware and contains: (i) a segment buffer providing 2$\times d$-wide weight fragments, (ii) a bit-error checker, (iii) a mixed-precision MAC unit, and (iv) a commit buffer with scoreboard interfaces that track segment readiness and pending corrections.

Segments are scheduled by readiness rather than strict order, making each lane resemble a narrow out-of-order pipeline. The scoreboard tracks whether a segment has passed error detection (valid bit = 0), is waiting for correction (valid bit = 1), or has been committed (no entry). This allows the lane to continue executing other ready segments even while others undergo multi-cycle correction. Final accumulation follows the in-order completion of weight columns, ensuring numerical correctness and matching the fixed page-level read address accumulation mode of the plane side.

A fly-weight register guarantees that each lane consumes exactly one segment per cycle without reshaping. In this design, the double-buffer structure also ensures that when a flying segment encounters an error, the MAC unit can immediately consume another weight segment. This is implemented via a bypass path between the fly-weight register and the PE lane. Given the extremely low probability of consecutive segment errors, it is reasonable to assume that another segment can be executed without delay. To avoid any potential loss of accuracy, once an OoO-ECDP enters this state, the faulty buffer is temporarily masked until the next cycle, when the data and corresponding check bit are transferred by the arbiter to the scoreboard. The segment is then automatically corrected. If the corrected weight segment matches the original, the scoreboard removes the entry directly.

This architecture allows the pipeline to sustain a full-cycle MAC rate even under non-zero RBER, while amortizing the correction logic across lanes. The datapath includes fused multiply-accumulate units supporting BF16 and INT8 execution, with optional widening for high-dynamic-range Transformer layers.

\subsection{Structured Prefetch and Pipeline Scheduling}

Because NAND reads are deterministic at page granularity, prefetching can be tightly coordinated with lane-level compute. NVLLM issues prefetches on a per-cluster, per-layer schedule, filling cluster FIFOs ahead of ERDPE consumption without generating additional random reads. The scheduler integrates with the OoO mechanism: segments with earlier deadlines are prefetched first, while segments experiencing correction stalls do not throttle the pipeline. This structure contrasts with DRAM-oriented prefetchers that must predict irregular access patterns; here, FFN weights are laid out contiguously across plane clusters, enabling deterministic, deadlock-free scheduling.

\subsection{End-to-End Dataflow}

As a 3D NAND-centric architecture, NVLLM coordinates data flow across the NAND array, NAND CMOS, NPU, and DRAM. To minimize energy-intensive transfers, recurrent weight accesses remain within either the 3D NAND or the NPU-DRAM subsystem; notably, Q/K/V/O weights are copied once into DRAM at initialization, eliminating all subsequent Flash-to-DRAM movement.

During prefill, the global scheduler distributes Q/K/V/O computation between NAND CMOS and the NPU based on their effective compute capabilities. The ERDPE performs in-Flash linear projections, after which the NPU fuses the $Q$, $K$, and $V$ matrices, executes attention, and stores $K$/$V$ in DRAM. Multi-head output projection follows an identical compute split, while FFN computation and the final output projection are both executed on the NAND side. The resulting logits are then transferred to the NPU for post-processing.

Decode follows a similar pipeline but processes a single token per step. At the early stage of decoding, attention is executed entirely on the NPU, while FFN computation is offloaded to the NAND side. As the KV cache grows, NPU-side attention latency becomes non-stationary, leading to imbalance in the shared Q/K/V/O path. NVLLM addresses this via a KV-cache-aware dynamic scheduler that adaptively adjusts the NAND-NPU compute split in real time using a lightweight latency estimator and a bitmap-based dispatcher. Algorithm~\ref{alg:kv_aware} illustrates a concrete instantiation of the proposed adaptive scheduling mechanism.
\begin{algorithm}
\caption{KV-Cache-Aware Scheduling}
\label{alg:kv_aware}
\begin{algorithmic}[1]

\Statex \textbf{Input:} Cycle increment $\Delta C$ (NPU cycles), NPU latency per column $C_\text{NPU}$, weight column size $u$, page buffer per plane cluster $P$, cluster-level bitmap $B^{(n)} \in \{0,1\}^{1 \times H}$
\Statex \textbf{Output:} Updated bitmap $B^{(n+1)}$
\Statex \textit{Executed at the end of each forward propagation}
\State $C_{\text{th}} \gets \lfloor P/u \rfloor \cdot C_\text{NPU}$
\If{$\Delta C \le C_{\text{th}}$} 
    \State \Return $B^{(n)}$
\EndIf
\State $k \gets \lceil \Delta C / C_{\text{th}} \rceil$
\State $B^{(n+1)} \gets B^{(n)},\ cnt \gets 0,\ i \gets H-1$
\While{$cnt < k$ \textbf{and} $i \ge 0$} 
    \If{$B^{(n)}[i] = 1$} 
        \State $B^{(n+1)}[i] \gets 0$
        \State $cnt \gets cnt+1$
    \EndIf
    \State $i \gets i-1$
\EndWhile
\State \Return $B^{(n+1)}$
\end{algorithmic}
\end{algorithm}
\section{Evaluation}

\subsection{Methodology}

\textbf{Hardware Configuration.} 
We model the 3D NAND Flash array at the plane level using the 3D-FPIM framework~\cite{lee3DFPIMExtremeEnergyEfficient2022b}, which captures key technology parameters such as cell and wordline geometries, parasitic capacitances, and timing characteristics validated against commercial products. The W2W stacking in NVLLM is reflected by adapting the layout hierarchy in area estimation, constrained by teardown analyses of consumer-grade NAND products~\cite{xtacking_vlsi}. LPDDR5X DRAM serves as shared memory for Q/K/V/O weights and the KV cache, with timing and energy modeled using Ramulator2~\cite{luo2023ramulator2} and DRAMPower~\cite{drampower}. A cycle-accurate C++ simulator integrates the 3D-FPIM NAND and DRAM behavioral models to evaluate system-level performance under diverse LLM workloads.

The NAND subsystem includes 32 planes organized into $2 \times 2$ clusters, each with a 16\,KiB page buffer, yielding a total capacity of 128\,GiB. Each plane occupies 3.07\,mm\textsuperscript{2} and exhibits a read latency of 5.12\,$\mu$s. We also evaluate scaled variants NVLLM-12C and NVLLM-16C to study architectural scalability.

\begin{table*}[t]
  \centering
  \fontsize{7.5}{8.5}\selectfont
  \setlength{\tabcolsep}{5pt}
  \renewcommand{\arraystretch}{1.16}
  \caption{Baseline Hardware Configuration Details}
  \label{table:baseline}
  \begin{tabular}{|c|c|c|c|c|c|c|c|}
    \hline
    \textbf{Architecture/System} & \multicolumn{3}{c|}{\textbf{GPU-Centric}} & \multicolumn{3}{c|}{\textbf{SSD-Like}} & \textbf{3D NAND-Centric} \\
    \hline
    \textbf{Design Name} & GPU-DRAM & GPU-SSD & GPU-Hybrid & Cambricon-LLM & AiF & AiF{-}{-} & NVLLM-16C \\
    \hline
    \textbf{Weight Placement} & DRAM & NAND Flash & NAND Flash + DRAM & \multicolumn{3}{c|}{NAND Flash} & NAND Flash + DRAM \\
    \hline
    \textbf{Quantization} & \multicolumn{7}{c|}{8bit} \\
    \hline
    \makecell{\textbf{Hardware} \\ \textbf{Configuration}} & 
    \multicolumn{3}{c|}{\begin{tabular}[c]{@{}c@{}}NVIDIA A800, 80GB GPU;  Intel Xeon \\Gold 6348 CPU;  512GiB DRAM, \\16$\times$32GiB DDR4; 5.2TB NVMe SSD\end{tabular}} & 
    \begin{tabular}[c]{@{}c@{}}8ch, 2chip/ch,\\ 2die/chip, 2plane/die,\\ 64 planes, 16KiB page\end{tabular} & 
    \multicolumn{2}{c|}{\begin{tabular}[c]{@{}c@{}}8ch, 2chip/ch,\\ 4plane/chip,\\ 64 planes, 16KiB page\end{tabular}} & 
    \begin{tabular}[c]{@{}c@{}}16 clusters, \\ 4 planes/cluster,\\ 64 planes, 16KiB page\end{tabular} \\
    \hline
  \end{tabular}
\end{table*}

\textbf{Benchmark Models.} 
We adopt the OPT series---OPT-1.3B, OPT-2.7B, OPT-6.7B, OPT-13B, and OPT-30B---as representative benchmarks covering a broad parameter range. To demonstrate NVLLM’s generality, we also evaluate LLaMA-family models. All models are quantized to INT8 for on-device inference. Random bit flips are injected into model weights to emulate realistic NAND RBER patterns, assessing robustness under non-ideal storage.

\textbf{Baselines and Metrics.} 
We compare NVLLM against conventional GPU-based inference and SSD-like in-Flash computation designs. GPU-centric baselines include GPU-DRAM and GPU-SSD, using an NVIDIA A800 GPU with DRAM or NAND Flash SSDs for model weights, implemented via FlexGen~\cite{sheng2023flexgen}. SSD-like baselines comprise Cambricon-LLM~\cite{yuCambriconLLMChipletBasedHybrid2024} and AiF~\cite{AiFAcceleratingOnDeviceLLMInference2025a}, which embed compute logic within Flash chips and coordinate multiple chips via SSD-style channels. NVLLM-16C is configured with 16 clusters and 64 planes for fair comparison with these designs. Table~\ref{table:baseline} summarizes the baseline configurations.

Performance metrics include tokens-per-second (TPS) for throughput, seconds per inference (s/inf) for end-to-end latency, and \si{\joule/token} for energy efficiency.

\subsection{Area and Power Estimation}

The NPU and NAND CMOS components of NVLLM were implemented in Verilog HDL and synthesized using Synopsys Design Compiler with TSMC 28\,nm technology. Table~\ref{table:Area} summarizes area and power breakdowns, reporting compute units and ECC logic separately. The ECC module achieves $2.0\times$ area and $1.3\times$ power reduction compared to a monolithic ECC design.

The NAND CMOS subsystem integrates SRAM blocks, including a 512\,KiB cache FIFO (reusing internal Flash cache resources), a 72\,KiB global buffer, and a 16\,KiB activation buffer. A lightweight RISC-V CPU coordinates register initialization and runtime operations. Overall, the NPU occupies 0.46\,mm\textsuperscript{2}, while the additional in-Flash logic occupies 2.69\,mm\textsuperscript{2}---accounting for only 2.7\% of the total NAND CMOS area.

\begin{table}[h!]
  \centering
  \fontsize{7.8}{8.5}\selectfont
  \renewcommand{\arraystretch}{1.14}
  \caption{Area and Power Overhead of Compute Core}
  \label{table:Area}
  \begin{tabular}{|c|c|c|c|}
    \hline
    \makecell{\parbox{1.6cm}{\centering \textbf{Component}}} & \textbf{Module} & \makecell{\parbox{1.4cm}{\centering \textbf{Area} ($\mu$m$^2$)}} & \makecell{\parbox{1.5cm}{\centering \textbf{Power} ($mW$)}} \\
    \hline
    \multirow{5}{*}{\textbf{NPU}} & SFU & 8,618 & 2.73 \\
    \cline{2-4}
                         & Dot-Product Unit & 144,712 & 170.40 \\
    \cline{2-4}
                         & SRAM & 304,217 & 67.00 \\
    \cline{2-4}
                         & Others & 1,767 & 0.02 \\
    \cline{2-4}
                         & NPU Total & 459,314 & 240.15 \\
    \hline
    \multirow{7}{*}{\makecell[c]{\textbf{NAND} \\ \textbf{CMOS}}} & RISC-V CPU & 685,284 & 2.76 \\
    \cline{2-4}
                         & Dot-Product Unit & 289,424 & 340.80 \\
    \cline{2-4}
                         & Detector (x8) & 82,256 & 159.69 \\
    \cline{2-4}
                         & Corrector (x8) & 323,608 & 107.66 \\
    \cline{2-4}
                         & SRAM & 1,292,922 & 284.75 \\
    \cline{2-4}
                         & Others & 18,089 & 0.02 \\
    \cline{2-4}
                         & NCW Total & 2,691,583 & 895.68 \\
    \hline
  \end{tabular}
\end{table}
\begin{table}[h!]
  \centering
  \fontsize{7.6}{8.5}\selectfont
  \renewcommand{\arraystretch}{1.06}
  \caption{NVLLM Scaling Configuration}
  \label{table:Confgurations}
  \begin{tabular}{|c|c|c|c|}
    \hline
    \makecell{\parbox{1.6cm}{\textbf{Configuration}}} & \makecell{\parbox{1.6cm}{\centering \textbf{NVLLM}}} & \makecell{\parbox{1.6cm}{\centering \textbf{NVLLM-12C}}} & \makecell{\parbox{1.6cm}{\centering \textbf{NVLLM-16C}}} \\
    \hline
    \textbf{NPU}	& \multicolumn{3}{c|}{4$\times$OoO-ECDP(w/o ECC), 2$\times$LPDDR5X, 6$\times$16Gb} \\
    \hline
    \textbf{OoO-ECDP} & 8 & 12 & 16 \\
    \hline
    \textbf{Plane Cluster} &	8 &	12 & 16 \\
    \hline
    \textbf{Plane} &	32 & 48 & 64 \\
    \hline
  \end{tabular}
 
\end{table}
\subsection{Throughput}

\begin{figure}[t]
  \centering
  \includegraphics[width=\linewidth]{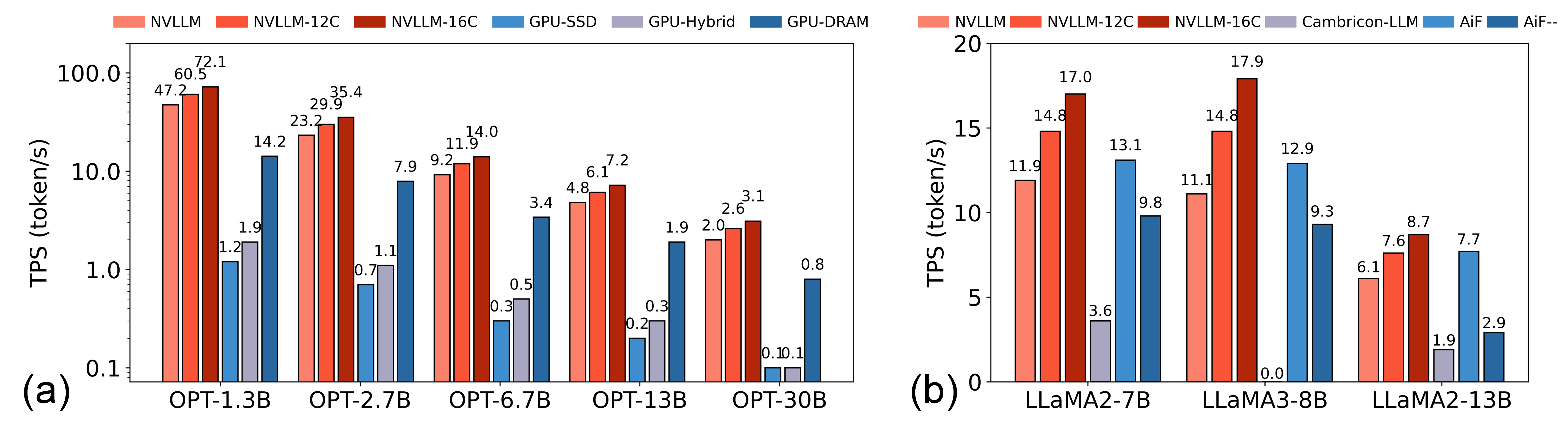}
  \vspace{-19pt} 
  \caption{Throughput comparison result.}
  \label{fig:fig6}  
\end{figure}

\textbf{Comparison with GPU-centric Baselines.} 
Although the NVIDIA A800 GPU delivers up to 624 TOPS of INT8 compute, generating one token of OPT-30B requires only 59.9 GOPs, leaving a substantial compute-utilization gap. Fig.~\ref{fig:fig6}(a) compares NVLLM, NVLLM-12C, and NVLLM-16C against three GPU-centric baselines. NVLLM operates its NAND CMOS at 350\,MHz and the NPU at 500\,MHz, delivering a total of 307--486\,GOPS. All decoding tests are conducted under a 64-token context.
Compared to GPU-SSD, NVLLM achieves $22.4\times$--$37.9\times$ speedup, leveraging multi-plane parallelism and internal 3D NAND bandwidth up to 100~GB/s, far exceeding typical PCIe Gen4 $\times$4 bandwidth limits of 8~GB/s. GPU-DRAM offers higher bandwidth than SSD but still falls short of NVLLM, yielding $>2.5\times$ lower throughput. Smaller models benefit more due to lower compute contribution relative to bandwidth.

\textbf{Comparison with SSD-like Architectures.} 
Fig.~\ref{fig:fig6}(b) compares NVLLM-16C with Cambricon-LLM, AiF, and AiF{-}{-}. For LLaMA2-7B, NVLLM-16C outperforms these designs by $4.7\times$, $1.3\times$, and $1.7\times$, respectively. Cambricon-LLM's limited throughput (3.6\,tokens/s) is due to shared Flash resources serving both in-Flash compute and NPU weight fetches. AiFSSD achieves 13.1\,tokens/s with 102.4~GB/s internal bandwidth; AiF{-}{-}, with reduced ECC/read optimizations, drops to 9.8\,tokens/s.

NVLLM’s advantage stems from hybrid-bonding enabling plane-level parallelism and higher internal bandwidth, and decoupled attention/FFN computation in NAND CMOS, allowing fully standalone inference without host-side coordination.

\subsection{Latency Analysis}

\begin{figure}[t]  
\centering
\includegraphics[width=\linewidth]{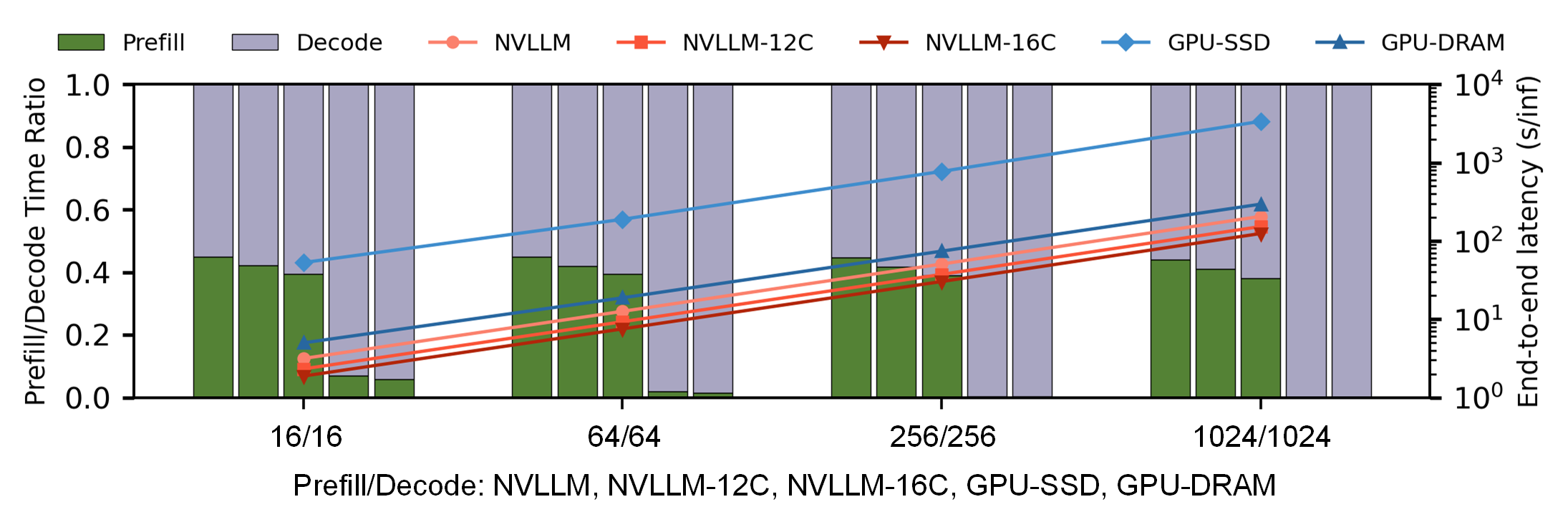}
\vspace{-15pt} 
\caption{End to end latency comparison result.}
\label{fig:fig7}  
\end{figure}

End-to-end latency of NVLLM is shown in Fig.~\ref{fig:fig7}, compared only to GPU-centric baselines due to missing prefill data for Cambricon-LLM and AiF. Prefill/decode token pairs range from 16/16 to 1024/1024. NVLLM distributes latency more evenly: prefill accounts for 44.1\%--45.0\% versus 0.1\%--6.9\% (GPU-SSD) and 0.15\%--5.84\% (GPU-DRAM).  

This reflects fundamental differences: GPUs are compute-rich, making prefill fast and decode bandwidth-limited; NVLLM decomposes GEMM/GEMV into dot-product primitives, with 3D NAND + LPDDR5X supplying sufficient bandwidth for FFN and attention, keeping the prefill phase compute-bound. Increasing plane clusters with OoO-ECDP units (Table~\ref{table:Confgurations}) further reduces prefill proportion. NVLLM-16C achieves 1.9\,s, 7.5\,s, 30.3\,s, and 124.3\,s per inference for 32, 128, 512, and 2048 tokens, up to $28.2\times$ faster than GPU-SSD and $2.7\times$ faster than GPU-DRAM.

\begin{figure}[t]  
\centering
\includegraphics[width=\linewidth]{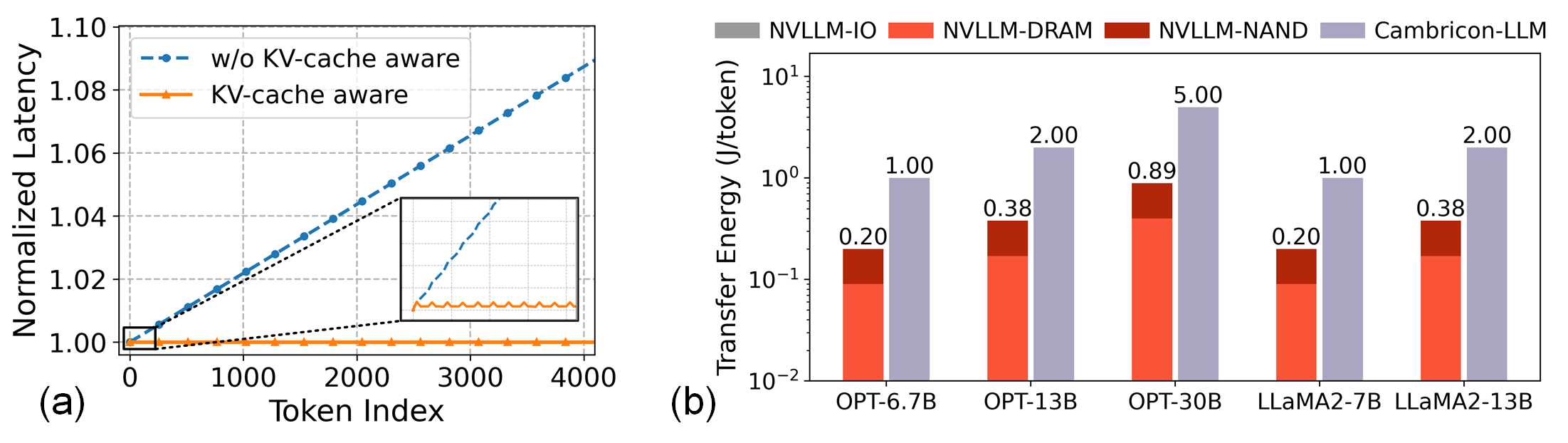}
\vspace{-13pt} 
\caption{(a) The effect of KV-cache-aware scheduling. (b) Energy comparison result.}
\label{fig:fig8}
\end{figure}

Furthermore, the KV-cache-aware scheduling mechanism enhances NVLLM’s latency benefits in extended decoding scenarios, which is validated by the ablation results shown in Fig.~\ref{fig:fig8}(a).
\subsection{Energy Efficiency}

NVLLM reduces data movement energy via decoupled attention/FFN design. Transfers occur along three paths: NAND array $\rightarrow$ NAND CMOS (NVLLM-NAND), and between NAND CMOS and the NPU (NVLLM-IO), and between the NPU and DRAM (NVLLM-DRAM). Most FFN weights remain within 3D NAND, and the decoupled architecture minimizes transfers between NAND CMOS and the NPU. Fig.~\ref{fig:fig8}(b) shows NVLLM's energy advantage over Cambricon-LLM. As model size increases, NVLLM's relative savings grow due to FFN-heavy workloads. NVLLM-IO is triggered only sparsely---primarily during layer transitions and the final output projection---and therefore contributes negligibly to the overall energy. Overall, NVLLM achieves $5.63\times$ reduction in data movement energy compared to Cambricon-LLM.

\section{Conclusion}

We present NVLLM, a 3D NAND-centric architecture for on-device LLM inference, co-designing NAND Flash, CMOS logic, NPU, and DRAM to maximize internal bandwidth and minimize data movement. By decoupling attention and FFN computation, leveraging multi-plane parallelism, and adopting dynamic KV-cache-aware scheduling, NVLLM achieves up to $37.9\times$ throughput improvement, $28.2\times$ latency reduction, and $5.6\times$ energy reduction over GPU- and SSD-based baselines. Our results demonstrate that carefully orchestrated near-data processing in 3D NAND enables efficient, scalable, and standalone LLM inference on resource-constrained devices.

\begin{acks}
This work was supported by the National Natural Science Foundation of China under Grant Nos. 92464204 and 92464302.
\end{acks}

\appendix

\section{Discussion: NVLLM Design Choices and Broader Implications}
\label{app:discussion}

This appendix discusses the design rationale of NVLLM, its intended deployment scenarios, and its broader implications for memory-system-driven AI workloads.

\subsection{Why a 3D NAND-Centric Architecture?}

NVLLM is motivated by a simple observation: as models continue to grow, the memory system increasingly determines the overall efficiency of edge AI inference. In this context, many architectural variants are foreseeable. One could imagine more aggressive forms of in-storage execution, deeper coupling between Flash and programmable accelerators, or broader mappings of Transformer operators into the NAND subsystem. NVLLM does not rule out these possibilities. Instead, it deliberately chooses a narrower and more concrete first step.

The design starts from constraints that are already present in practical 3D NAND systems, including page-granular reads, limited CMOS logic budget, raw read errors, non-negligible access latency, existing storage interfaces, and the need to preserve conventional read/write behavior. Under these constraints, the key question is not how many operators can be theoretically mapped into Flash, but which part of the workload exposes a stable architectural bottleneck that current or near-term technology can realistically address.

More broadly, NVLLM follows a problem-grounded view of architecture research. Rather than treating architectural freedom itself as the starting point, the design starts from constraints that are visible in current and near-term technology, including device behavior, process limits, storage interfaces, reliability, and system integration cost. From this perspective, the role of architecture is to identify which emerging bottlenecks have become first-order system problems under these constraints, and what existing technologies must support in order to make such solutions practical. NVLLM is developed in this spirit: it treats 3D NAND Flash as a constrained but increasingly relevant execution substrate, and focuses on the part of LLM inference where model capacity, data movement, and edge-system resources already meet as an architectural problem.

NVLLM answers this question by targeting the weight-dominated and regular portion of LLM inference. This choice reflects a design philosophy rather than a lack of alternatives. The goal is to first solve the architectural problem that emerges when large model weights reside in high-capacity non-volatile storage but must be repeatedly moved through a conventional storage-memory-compute hierarchy. By focusing on this problem, NVLLM provides a concrete point in the design space for 3D NAND-centric inference.

\subsection{Rationale Behind the FFN/Attention Partition}

The FFN/attention partition in NVLLM follows directly from this philosophy. FFN layers account for a large fraction of model parameters and exhibit regular weight-streaming behavior. Their access pattern naturally aligns with the page-level organization and plane-level parallelism of 3D NAND Flash. Executing FFN computation close to Flash therefore reduces repeated weight movement while keeping the in-Flash logic simple enough to respect the CMOS area and reliability constraints of NAND devices.

Attention and KV-cache management have different characteristics. Their behavior depends on sequence length, cache layout, sparsity, retrieval, and runtime scheduling policies. NVLLM keeps these components in the xPU--DRAM domain because they benefit from flexible control, random access, and fast adaptation to algorithmic changes. This is not a claim that dynamic operators are outside the scope of future storage-integrated systems. Rather, it is a conservative partition that allows NVLLM to support an evolving workload landscape without forcing every emerging operator into a fixed Flash-side execution model.

In this sense, NVLLM is designed with workload diversity in mind. As retrieval augmentation, sparse attention, speculative decoding, KV-cache optimization, and dynamic routing become more common~\cite{dynWork_ref1,dynWork_ref2,dynWork_ref3,dynWork_ref4}, the system can still preserve a stable in-Flash path for regular model-weight access while leaving dynamic runtime behavior to programmable compute and DRAM. The architecture therefore separates what is structurally stable from what is algorithmically fluid.

\subsection{Target Deployment Scenarios}

\begin{figure}[t]
  \centering
  \includegraphics[width=\linewidth]{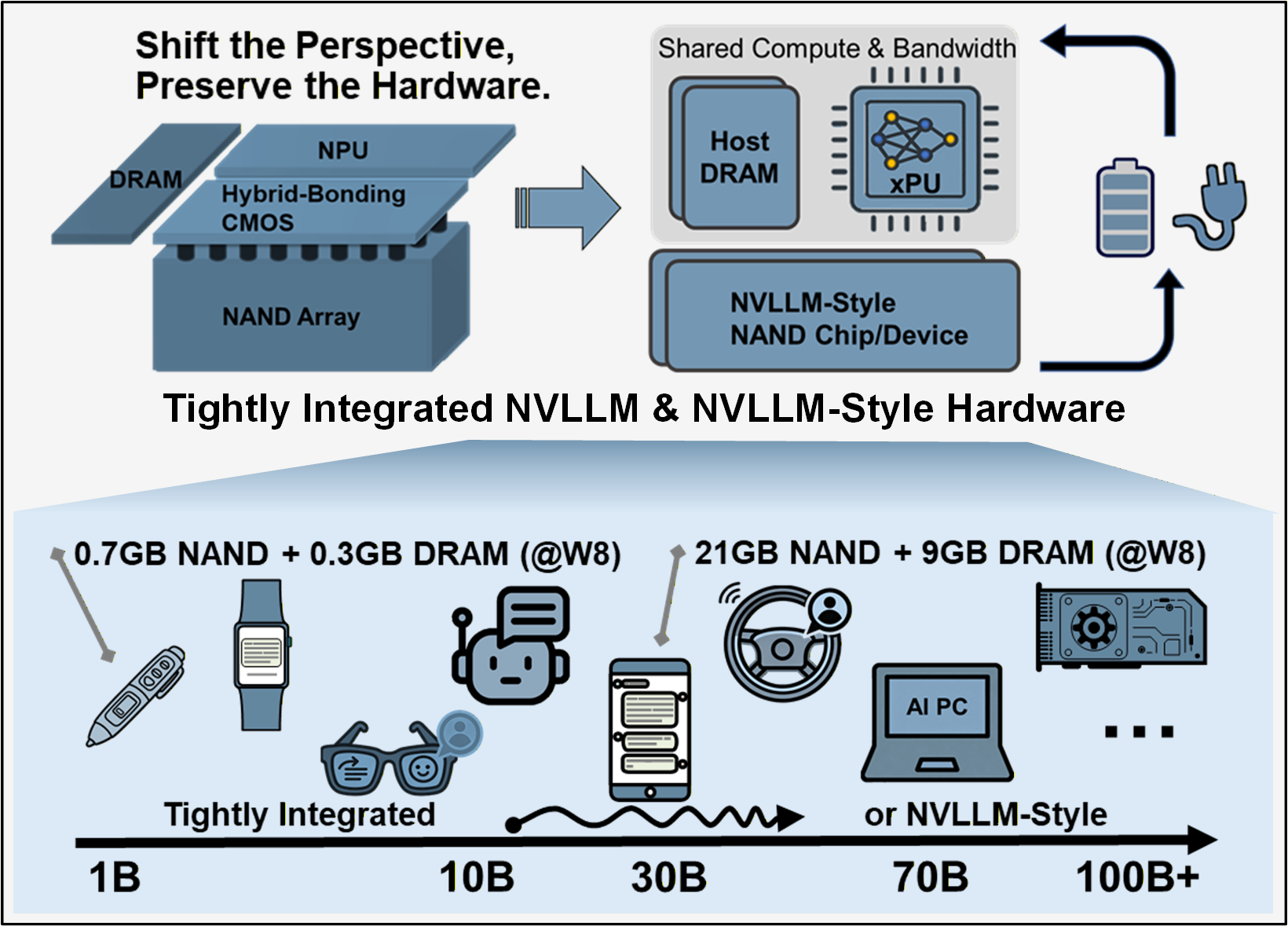}
  \caption{Illustrative deployment scenario of NVLLM in an edge platform. The regular FFN weight-streaming path is served by the 3D NAND-centric execution substrate, while dynamic attention, KV-cache management, and system-level tasks remain on the xPU--DRAM path.}
  \label{fig:nvllm_deployment}
  \vspace{-3pt}
\end{figure}

NVLLM is primarily intended for on-device LLM inference on platforms where model capacity exceeds available DRAM capacity. Representative examples include mobile devices, AI PCs, embedded assistants, privacy-sensitive offline applications, and embodied AI. These systems already integrate NAND Flash, DRAM, and CPU/NPU subsystems, making them natural candidates for a 3D NAND-centric execution path.

As illustrated in Fig.~\ref{fig:nvllm_deployment}, a practical advantage of NVLLM is that it preserves the native NAND read/write path and the external storage interface. The Flash subsystem can still serve as conventional storage, while selected inference phases use the NAND-side compute path. This property is important for real edge platforms, where an accelerator architecture must coexist with the operating system, file system, storage controller, and other application workloads.

Although NVLLM is evaluated in the context of LLM inference, the underlying principle is broader. Other emerging AI workloads, such as diffusion transformers (DiTs)~\cite{dit_ref1}, vision-language-action (VLA) models~\cite{vla_ref1}, and multimodal foundation models~\cite{mfm_ref1}, may also contain memory-bound phases when deployed under tight edge memory and power budgets. NVLLM does not imply that these workloads should be mapped wholesale into Flash. Instead, it suggests a methodology: when a workload contains large, regular, repeatedly accessed model states alongside dynamic runtime computation, the regular path may be a candidate for near-Flash execution, while the dynamic path remains in programmable memory and compute.

\subsection{Outlook: Toward 3D NAND-Native AI Systems}

Looking forward, AI-oriented 3D NAND should be optimized not only for storage density, but also for its role in model execution. Capacity, plane-level parallelism, page-buffer organization, read latency, ECC behavior, storage interfaces, and CMOS logic budget should be considered jointly. These factors determine whether high-capacity non-volatile memory can become an effective execution substrate rather than only a passive repository for model weights.

NVLLM is therefore best viewed as an early architectural step rather than a final form of 3D NAND-native AI hardware. Its contribution is to identify a near-term, technology-grounded opportunity: execute the stable, memory-bound portion of inference where the model weights reside, while preserving flexibility for dynamic and rapidly evolving operators. More broadly, NVLLM points to a design direction in which storage, memory, and compute are co-designed around the actual bottlenecks of large AI models, rather than treated as separate layers connected only by data movement.

\bibliographystyle{ACM-Reference-Format}
\bibliography{hmb}

\end{document}